\documentclass[aps,prl,onecolumn,superscriptaddress,10pt,notitlepage]{revtex4-1}
\usepackage{graphicx}
\usepackage{latexsym}
\usepackage{color}
\usepackage{amsbsy,amsmath,amssymb}

\begin{document}

\title{A spin dynamics approach to solitonics}

\author{Konstantinos Koumpouras}
\email[Corresponding author. ]{konstantinos.koumpouras@physics.uu.se}
\affiliation{Department of Physics and Astronomy, Uppsala University, P.O. Box 516, 751 20 Uppsala, Sweden}
\author{Anders Bergman}
\affiliation{Department of Physics and Astronomy, Uppsala University, P.O. Box 516, 751 20 Uppsala, Sweden}
\author{Olle Eriksson}
\affiliation{Department of Physics and Astronomy, Uppsala University, P.O. Box 516, 751 20 Uppsala, Sweden}
\author{Dmitry Yudin}
\affiliation{ITMO University, Saint Petersburg 197101, Russia}

\begin{abstract}
In magnetic materials a variety of non-collinear ground state configurations may emerge as a result of competition among exchange, anisotropy, and dipole-dipole interaction, yielding magnetic states far more complex than those of homogenous ferromagnets. Of particular interest in this study are particle-like configurations. These particle-like states, e.g., magnetic solitons, skyrmions, or domain walls, form a spatially localised clot of magnetic energy. In this paper we address topologically protected magnetic solitons and explore concepts that potentially might be relevant for logical operations and/or information storage in the rapidly advancing filed of solitonics (and skyrmionics). An ability to easily create, address, and manipulate such structures is among the prerequisite forming a basis of ``-onics technology'', and is investigated in detail here using numerical and analytical tools.
\end{abstract}

\maketitle
\vspace{1.0cm}
Low-dimensional magnetic systems remain a challenge, representing an exciting part of modern condensed matter physics. Tremendous progress in fabrication and characterization of magnetic nanostructures \cite{Magnano} allows to approach experimental investigations of nonlinear phenomena, which previously were considered to form a purely theoretical research area. From both fundamental scientific and technological perspectives, consistent studies of controllable motion of spin textures, that emerge in magnetic materials \cite{Kosevich}, pose many of the most interesting scientific questions. Depending on distance between neighbour spins, hybridization effects and details of the electronic structure as well as the crystalline symmetry, various magnetic configurations may emerge. These configurations range from collinear ferromagnetic and antiferromagnetic states, to non-collinear helimagnets and more complicated textures. In chiral magnets, where magnetic spiral ordering with a given sense of rotation is present, a non-collinear configuration may result from the competition between symmetric Heisenberg exchange and anisotropic antisymmetric Dzyaloshinskii-Moriya (DM) coupling. Indeed, magnetocrystalline structures which support spiral ground states are characterized by the lack of inversion symmetry and, as a consequence, antisymmetric exchange DM interaction. The latter takes place e.g. in metallic MnSi \cite{Grigoriev}, insulating CuB$_2$O$_4$ \cite{Petrakovskii}, and semiconducting Cr$_{1/3}$NbS$_2$ \cite{Moriya}.

Of particular interest for magnetic nano-objects are magnetic structures that are localized in space, i.e. solitons. An example of this is the domain wall, where a lot of knowledge has been obtained, both as concerns static \cite{Cervera} and dynamic \cite{Shibata} properties. Other magnetic solitons, like chiral spin-spirals and skyrmions \cite{Bogdanov89,Fert,Jiang} have remained less explored, so far. Interestingly, the balance between Heisenberg exchange and DM interaction allows for nontrivial topological spin-textures to exist as a ground state configuration of ferromagnets, that may develop towards two-dimensional chiral modulations, i.e. skyrmions. The skyrmion spin-configuration, with a fixed azimuthal angle, influenced by the underlying crystalline structure, might be stabilized in a non-centrosymmetric magnet \cite{Man}. Such a topological soliton may be stable both physically and topologically, i.e. it has a smooth defect-free core with a definite diameter that is characterized by material parameters only. Recently, the first experimental observation of a skyrmionic texture in a nano-sized, thin film structure were reported \cite{Nagaosa10}, for FeGe and Fe$_{0.5}$Co$_{0.5}$Si. Experiments reported a stable skyrmion phase in a wide temperature window, well below the Curie point, by means of Lorentz transmission electron microscopy. In another, recent, work a soliton phase was reported, experimentally, in an antiferromagnetic chain compound, KCuGaF$_6$ \cite{PhysRevB.92.174412}.

In general, the class of quasi-one-dimensional magnetics includes a variety of systems ranging from metal-organic complexes, consisting of weakly coupled spin chains, to helimagnets, in which magnetisation varies in one spatial dimension. From practical aspects, the most interesting class of compounds is the one where there exists a possibility to manipulate non-trivial topological structures by means of an external field or current.  
Furthermore, chiral magnets attract a lot of attention not only for being interesting from a fundamental viewpoint but also as a promising candidate for future spintronic devices. As mentioned, the compound Cr$_{1/3}$NbS$_2$ \cite{Moriya} is of interest, since a chiral soliton lattice has been proposed for its magnetic structure \cite{Ovchinnikov2012}. Its hexagonal structure is built up from NbS$_2$ layers intercalated by Cr ions. There are three basic magnetic interactions present in this structure: the ferromagnetic exchange within the Cr layers $J_\perp$, another ferromagnetic interaction between such Cr layers, $J_\parallel$, and the DM interaction between the Cr ions. The latter two interactions belong to the two intercalating layers separated by NbS$_2$ \cite{Ishikawa,Dyadkin}, and the competition between them forms a helical structure.  

There are two ways to drive the motion of a chiral soliton lattice: either by injecting a spin-polarised current into a sample (incoherent method), or by twisting the phase angle of the magnetic order parameter which directly couples to a magnetic field (coherent method). For example, one may envision the latter scenario using a rotating magnetic field, that is applied to one end of the sample, and that is sufficiently strong to orient the magnetisation parallel to it. Because of the stiffness of the spin system, the spin rotation at one end is transmitted to the other end of the sample, even if it is not subject to the direct effect of the applied rotating magnetic field. Transmission of the torque through the sample hence presents a spin current. The incoherent current injection method to drive the sliding dynamics has already been proposed \cite{Ovchinnikov2010}, whereas chiral helimagnets are promising candidates to realise the coherent method \cite{Ovchinnikov2012b}. It has been pointed out \cite{Ovchinnikov2008a,Ovchinnikov2008b} that once a sliding motion of helimagnets is triggered, the soliton lattice maintains its persistent motion assisted by a generation of inertial mass. Another observable consequence of the coherent motion is the appearance of a spin motive force \cite{Maekawa}, when the time dependence of the longitudinal magnetic field manifests itself in the temporal regime of the spin motive force. Remarkably, the chiral soliton lattice is a macroscopically ordered object, which contains macroscopic amounts of magnetic solitons (kinks). Due to the very large number of solitons, the spin motive force of this lattice is expected to be strongly amplified as compared with that caused by a single magnetic domain wall in a ferromagnet \cite{Ovchinnikov2008a,Ovchinnikov2008b}.

In this paper we present a detailed and systematic analysis of a quasi-one-dimensional chiral soliton lattice, e.g. as has been recently observed in Cr$_{1/3}$NbS$_2$ \cite{Ovchinnikov2012}, and the soliton-soliton interaction. We combine our numerical findings, obtained from atomistic spin dynamics simulations, with analytical results, performed for a one-dimensional helical structure which is known to support a soliton lattice as a ground state, when an external magnetic field perpendicular to helical axis is applied. Interestingly, in contrast to a uniformly polarized ground state configuration the formation and propagation of solitons on top of a kink crystal leads to a macroscopic translation of magnetic texture which should be experimentally detectable. Numerical results are found to be in good agreement with the analytical theory and this allows us to touch upon nonlinear dynamics of solitons and spin waves in spin-spiral structures, opening thus a way for solitonics.

\section{Results}
\noindent{\bf Analytical model.} Detailed theoretical study of nonlinear magnetization dynamics is rarely to be accomplished in closed analytical form, however before we present the results of our numerical simulations we elaborate a simple toy model which allows us to include all relevant interactions and make predictions which will be shown to be qualitatively correct. We examine the behaviour of a quasi-one-dimensional ferromagnet with lack of inversion symmetry located along $z-$axis placed in the transverse magnetic field $\mathbf{H}=\left(H,\,0,\,0\right)$: The Hamiltonian (per a unit volume) of such a system 

\begin{equation}\label{magnen}
\tilde{h}=h_\mathrm{ex}+h_\mathrm{DM}+h_\mathrm{an}+h_\mathrm{Z}
\end{equation}

\vspace{0.5cm}
\noindent is known to incorporate the terms due to exchange $h_\mathrm{ex}=A\left(\partial_z\mathbf{M}\right)^2/2$, Dzyaloshinskii-Moriya $h_\mathrm{DM}=D\mathbf{M}\cdot\left(\nabla\times\mathbf{M}\right)$, and Zeeman $h_\mathrm{Z}=-\mathbf{M}\cdot\mathbf{H}$ interaction. We have also included $h_\mathrm{an}=BM_z^2/2$, a component which stands for easy-plane ($xy-$plane) anisotropy in the expression (\ref{magnen}). In the following we assume the magnetisation is determined by a vector field $\mathbf{M}(z,t)$, so that $\mathbf{M}^2=M_0^2=\mathrm{const}$, while $z$ and $t$ are coordinates of space and time; the parameters $A$, $D$, and $B$ represent exchange and DM couplings, and magnetic anisotropy respectively. In the absence of an external field DM favours the formation of a spiral structure along the $z-$axis, which is incommensurate with the underlying crystalline structure, whereas the magnetic field $\mathbf{H}$, applied in the $x-$direction, tends to align the spins in this direction. The delicate interplay between these two interactions may lead to the formation of elongated domains of width $L$ (see below in the text) in which magnetisation is uniform. There exists a critical field $H_c$, where as long as $H>H_c$ the system is collinear, ferromagnetic. 

The vector field $\mathbf{M}(z,t)$ allows a parametrisation with polar $\Phi\equiv\Phi(z,t)$ and azimuthal $\Theta\equiv\Theta(z,t)$ angles: $\mathbf{M}(z,t)=M_0\left(\sin\Theta\cos\Phi,\sin\Theta\sin\Phi,\cos\Theta\right)$. There are many magnetic materials where the magnetic anisotropy is very high, so that $H/M_0\leq D^2/A\ll B$ \cite{Borisov2009}. For such systems one can use $\Theta=\pi/2+\theta$, where $\theta\ll 1$. If this is the case, the Landau-Lifshitz equation, combined with the Hamiltonian in Eqn.\ref{magnen}, leads to the following equations for the leading order terms for $\theta$:

\begin{equation}\label{LL1}
\theta=\frac{\Phi_t}{\gamma M_0B},
\end{equation}

\begin{equation}\label{LL2}
-\frac{1}{\gamma M_0}\theta_t+A\Phi_{zz}-\frac{H}{M_0}\sin\Theta=0,
\end{equation}

\noindent where $\gamma$ is the gyromagnetic ratio. Note that substituting (\ref{LL1}) into (\ref{LL2}) and introducing dimensionless variables $\tilde{t}=\gamma t\sqrt{M_0BH}$ and $\tilde{z}=z\sqrt{H/(AM_0)}$ one arrives at the celebrated sine-Gordon model (see, e.g. \cite{Borisov,BorKis}):

\begin{equation}\label{sing}
\Phi_{\tilde{t}\tilde{t}}-\Phi_{\tilde{z}\tilde{z}}+\sin\Phi=0.
\end{equation}

\vspace{0.5cm}
\noindent With the help of (\ref{LL1}) we can rewrite energy density (\ref{magnen}) as follows:

\begin{equation}\label{magnen1}
h=\dfrac{\tilde{h}}{M_0H}=\dfrac{1}{2}\left[\left(\partial_{\tilde{z}}\Phi\right)^2+\left(\partial_{\tilde{t}}\Phi\right)^2\right]+\tilde{q}\partial_{\tilde{z}}\Phi+\left(1-\cos\Phi\right),
\end{equation}

\vspace{0.5cm}
\noindent where the pitch vector $\tilde{q}=D\sqrt{M_0/(AH)}$ coincides with its standard notation if we switch back to $z$ and $t$, i.e. $q=\tilde{q}\sqrt{H/(AM_0)}=D/A$. Depending on $\tilde{q}$, stationary solutions to Eqn.\ref{sing} are either a ferromagnetic state $\Phi\equiv0$ (mod $2\pi$) or a spiral structure $\Phi\equiv\varphi_0(z)=\pi-2\mathrm{am}\left(u,n\right)$, with $u=z/n$. Note that $\mathrm{am}(u,n)$ is the amplitude of an elliptical integral, and $n$ is the modulus of the elliptic function. It should be noted that $n$ can be thought of as a parameter of the theory and should be determined by minimising the total energy (\ref{magnen1}) of the system, resulting in

\begin{equation}\label{min}
\pi nq-4E(n)=0.
\end{equation}

\vspace{0.5cm}
\noindent Equation (\ref{min}) does not have any solution as long as $q<4/\pi$, so that the ground state is ferromagnetic, whereas for $q>4/\pi$ the system is in a spiral state, i.e. $\varphi_0(z)$ changes by $2\pi$ with period $L=2nK(n)$ [here we have used elliptical integrals of the first $K(n)$ and second $E(n)$ kind, respectively]. In the latter case, the minimisation procedure leads to

\begin{equation}\label{kink}
\partial_z\varphi_0(z)=-\frac{\pi}{K(n^\prime)}\sum\limits_{p=-\infty}^\infty\mathrm{sech}\left[\frac{\pi}{2K(n^\prime)}\Big(\frac{z}{n}-2pK(n)\Big)\right],
\end{equation}

\begin{figure}
\includegraphics[scale=2.7]{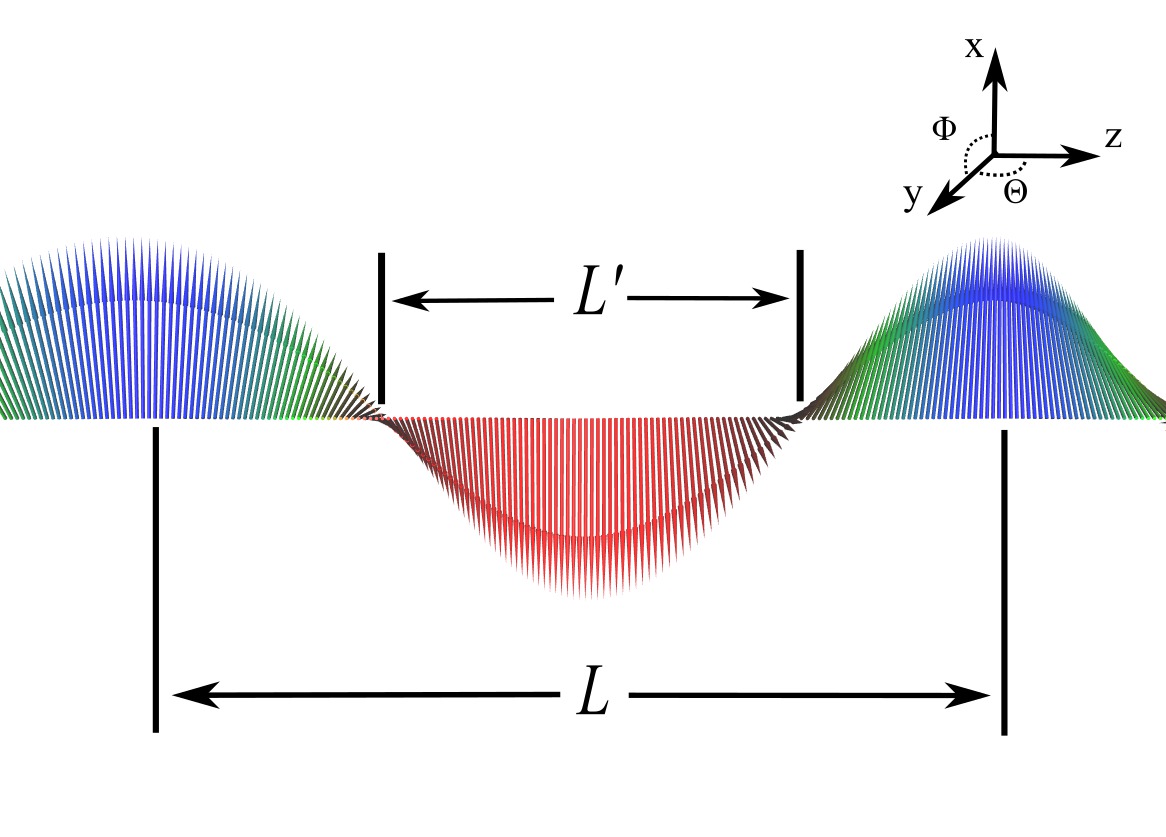}
\caption{\label{lattice}\textbf{Schematic visualisation of the soliton (this represents a zoomed-in picture of the system).} The direction of the external magnetic field is on $x−$-axis, the direction of the blue magnetic moments are parallel to it and the moments of the red region are antiparallel. $L$ stands for the period of the soliton and $L^\prime$ is the centre of the period. All the atomic magnetic moments are in $x-y$ plane}
\end{figure}

\vspace{0.5cm}
\noindent where $n^\prime=\sqrt{1-n^2}$. In general, $\varphi_0(z)$ changes in a tiny region around $z_p=2npK(n)$ with a characteristic lengthscale $l=2nK(n^\prime)/\pi$, so that $\varphi_0(z)$ can be approximated with a constant over large regions of space. Therefore, the ground state $\varphi_0(z)$ represents a lattice of $2\pi-$kinks separated by elongated regions $L$ (an example is shown in Fig.~\ref{lattice}).

As we have just seen, under certain physically motivated conditions the Landau-Lifshitz equation for magnetically ordered systems can be reduced to the well-known sine-Gordon problem, equation (\ref{sing}), where $\Phi$ defines magnetic order in a structure. All the solutions to equation (\ref{sing}) with uniform asymptotic behaviour at $z\rightarrow\pm\infty$ have been analysed with the inverse scattering method \cite{InvScat}. Interestingly, equation (\ref{kink}) with nontrivial ground state configuration can not be integrated out with standard methods. It is also reasonable to expect that the propagation of a soliton throughout the lattice of kinks differs from that with a homogeneous magnetisation \cite{BorKis}. The sine-Gordon equation (\ref{sing}) can be solved by means of B\"acklund auto-transformations leading in general to an $n$-solitonic solution  \cite{Backlund}. These transformations are known to link one solution to another, often linear, equation. It is noteworthy that the obtained solutions determine nonlinear superposition principle and in general allow to derive an infinite family of exact solutions, using straightforward algebra only. 

The formation of solitons is associated with local translation of the magnetic structure by $0<n\Delta<2L$. One can classify the general solution to belong to one of the following groups, discussed below. First we consider a structure characterized by the following asymptotic behaviour:

\begin{equation}\label{bc1}
\Phi(z,t)=\left\lbrace\begin{array}{cc}
\varphi_0(z), & z\rightarrow\infty \\
2\pi\sigma+\varphi_0(z+n\Delta), & z\rightarrow-\infty
\end{array}\right.
\end{equation}

\vspace{0.5cm}
\noindent where $\sigma=\pm1$: at $\sigma=1$ an extra kink of the same chirality emerges and the background structure (a lattice of $2\pi-$kinks) is shifted by $n\Delta$ along the helical axis. Note that this shift is smaller than a period of the kink crystal $L$. At $\sigma=-1$ the kink crystal grows by a period and an extra kink of the opposite chirality appears. Provided the expression (\ref{bc1}) takes place, one can expect the existence of an odd number of domain walls with mutually opposite chirality and an arbitrary number of breathers, which can be thought of as pulsatile soliton, i.e. a bound state of two kinks with opposite chiralities. The magnetic structure in this case allows propagation of spin waves which does not lead to the local translation, i.e. spin waves do not affect the nontrivial background structure. 

Another scenario is realized for the following asymptotic behaviour:

\begin{equation}\label{bc2}
\Phi(z,t)=\left\lbrace\begin{array}{cc}
\varphi_0(z), & z\rightarrow\infty \\
\varphi_0(z+n\Delta), & z\rightarrow-\infty
\end{array}\right.
\end{equation}

\vspace{0.5cm}
\noindent which allows an even number of kinks with mutually opposite chirality and an arbitrary number of breathers, while $n\Delta$ stands for the local translation due to soliton formation. To convince oneself that the expressions (\ref{bc1}) and (\ref{bc2}) do make physical sense one in general maps nonlinear differential equation to an equivalent system of differential equations (in mathematics this strategy is known as Lax transformation). For the sine-Gordon model (\ref{sing}) one can construct a solution which corresponds to translations of the background structure (see, e.g., Ref.\cite{BorKis}). When the analytical solution has been determined one can observe (see Supplementary Movies) that the formation of a soliton is associated with local translations of a lattice of $2\pi-$kinks which needs to be taken into account via asymptotic behavior (\ref{bc1}) and (\ref{bc2}). In fact, a newly created soliton interacts with surrounding kinks pushing them away.

Remarkably, in each case the background helical structure readjusts itself to counteract the formation of a soliton. This causes the soliton to move through a kink crystal resulting in local translation of the spiral structure by $n\Delta$ in a direction opposite to that in which the soliton travels. Interestingly, the scalar field $\Phi$ fluctuates when the soliton interacts with kinks of the background structure. In the vicinity of a commensurate-incommensurate phase transition ($n\rightarrow 1$) $\Delta\approx K(n)$. The collision of two solitons with opposite chiralities leads to partial annihilation and local magnetisation fluctuations determined by $\Delta$. Summarising, a wide class of helical magnets with inhomogeneous ground state configuration (spin spiral) promotes propagation of multisolitonic structure, so that it shifts the spin spiral slightly past the original position. This finding is as we will see in the following section, in good agreement with numerical simulations utilising an atomistic, discrete model. 

\vspace{0.5cm}
\noindent{\bf Numerical simulations} Meaningful information on the nature of the ground and excited states can be extracted from studying magnetisation dynamics by numerical methods. This is typically done by finding solutions of phenomenological Landau-Lifshitz equation equipped with Gilbert term \cite{Brataas}, that stands for dissipation, and a Slonczewski contribution responsible for current-driven phenomena. Both Gilbert damping and spin-transfer torque have to be treated on equal footing and can be microscopically derived \cite{Duine}; while spin-polarised electronic structure calculations, may be used to evaluate relevant parameters of an effective spin-Hamiltonian \cite{LKAG}. Hence exchange and DM interaction may be evaluated and used in an atomistic description of the spin-system, where also a dynamical description is possible. 

In the present study, we are interested in the current induced soliton dynamics in a one-dimensional lattice. Under the influence of an external magnetic field the spin structure is expected to change into kink crystal according to equation (\ref{kink}), as discussed above.
To study the motion of the solitons in 1D structure in the presence of a spin polarised current, we have performed atomistic spin dynamics simulations based on the Landau-Lifshitz-Gilbert (LLG) equation with additional terms to describe the spin transfer torque effect.
The investigation of dynamics of the solitons introduced in the system is performed in terms of atomistic spin dynamics method \cite{PhysRevB.54.1019}, as implemented in the UppASD package \cite{0953-8984-20-31-315203} (see Method).

\vspace{0.5cm}
\noindent{\bf Stabilisation and soliton structure.} Using models which are defined in Methods (equations (\ref{latham}) -- (\ref{LaLi})), we now turn to the task of describing the microscopic structure of the solitons considered earlier, as they result from numerical simulation based on an atomistic approach. In order to describe a soliton structure from the LLG simulations, we consider a one-dimensional chiral magnet with 1000 magnetic moments along the $z-$direction with periodic boundaries conditions. In the system the Heisenberg ferromagnetic exchange interactions and the DM interactions are considered between nearest neighbours only, so that in continuous limit the Hamiltonian (\ref{magnen}) is recovered. In the first calculation the strength of the ferromagnetic exchange interaction is $J=1.88$ mRy and the strength of the DM interaction is $D=0.08$ mRy. The magnetic moments are treated as classical vectors with fixed length and the magnitude 1 $\mu_B$, and values of the  dimensionless damping parameter were set to $\alpha=0.1$ or $\alpha=0.01$. As discussed below, the choise of damping parameter influenced the results only marginally.

\begin{figure}
\includegraphics[scale=0.40]{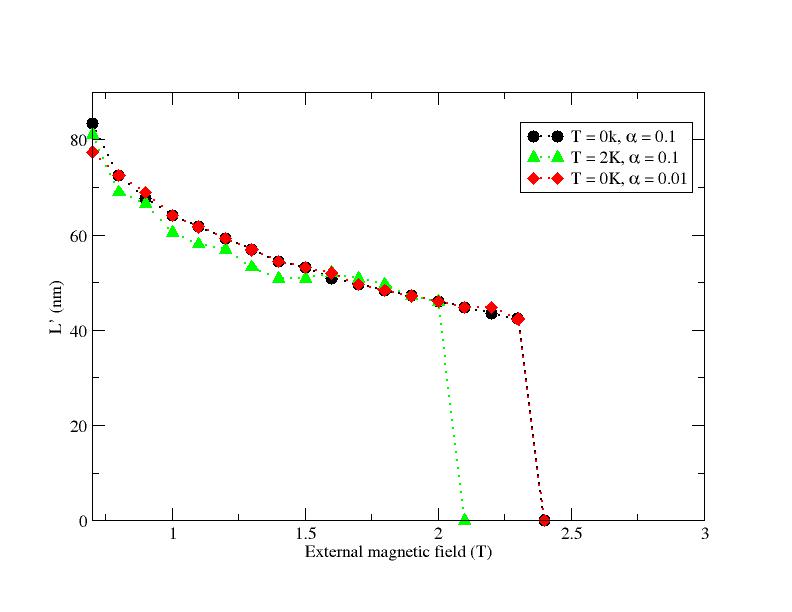}
\caption{\label{magnfield}\textbf{Size of the centre of one soliton ($L^\prime$) as a function of the external magnetic field.} This calculations are performed with parameters $J=1.88$ mRy, $D=0.08$ mRy and $D/J=0.042$ fot two different temperatures (0 and 2 K) and different values of damping (0.1 and 0.01). The dotted lines are used as guides for the size.}
\end{figure}

The spin configuration in Fig.~\ref{lattice} is stabilised by applying an external magnetic field $B=0.1$ T in the $x-$direction at zero temperature, and the period of the soliton $L$ is 174 nm. The size is determined by the $D/J$ ratio and the strength of the applied external magnetic field, in full agreement with our analytical findings described in equation (\ref{kink}), which gives a period of 177.8 nm. In this work the period of the soliton $L$ is determined by the distance between the magnetic moments with $m_x=1$ (centre of the blue domains) and $L'$ is determined by the distance between the magnetic moments at the edges of the red region with $m_x=0$ (see Fig.~\ref{lattice}). By performing atomistic spin dynamics simulations we can keep track of the characteristic size of the soliton's centre: It is known that on approaching the commensurate-incommensurate phase transition it collapses. The results of our simulations are shown in Fig.~\ref{magnfield} for two choise of damping parameter ($\alpha$ = 0.1 and $\alpha$ = 0.01) at zero temperature. For finite temperature (T = 2K), we only used one value of the damping parameter ($\alpha$ = 0.1). One can easily see that the systems with different damping are almost identical and upon increasing the external field, the size of the solitons decreases almost linearly with respect to the magnitude of the applied field. For the system which is held at finite temperature we observe that fluctuations are more pronoucned, but it follows the tendency of the zero temperature simulations. For sufficiently large fields, the external field dominates over the internal exchange interactions and as a result, the soliton lattice breaks down and we observe a phase transition to a commensurate ferromagnetic ordering. 

The data in Fig.~\ref{magnfield} show that the soliton is stable also at elevated temperatures, albeit with slightly different characteristics compared to the soliton at T=0 K. In order to study the possibility of stabilizing solutions depicted in Fig.~\ref{lattice} at even higher temperatures we note that in agreement with theoretical predictions, increased values of 
$J$ and $D$ should not influence the size of the soliton, as long as the ratio $D/J$ ratio is fixed. For this reason we increased the value of $J$ to 3.76 mRy and $D$ to 0.16 mRy, and investigated the stability as a function of applied field and temperature. We find that the soliton is stable at temperatures at least up to 10 K, and that at this temperature it takes an applied field of $B_{ex} = 3.2$ T to destabilize the soliton. Parameters of exchange and DM interaction of this strength are not atypical \cite{Kubler}, and hence this exercise shows that the soliton proposed here should be possible to study experimentally, not only at sub-Kelvin temperatures. The period of spiral structure can be tuned by adjusting the external magnetic field. Furthermore, our simulations fit well with the experimental results on Cr$_{1/3}$NbS$_2$, where magneto-optical measurements were made in the presence of external field, perpendicular to the spiral axis \cite{Ovchinnikov2012}. These experiments are furthermore in line with the analytical results presented here.

\begin{figure}
\includegraphics[scale=0.40]{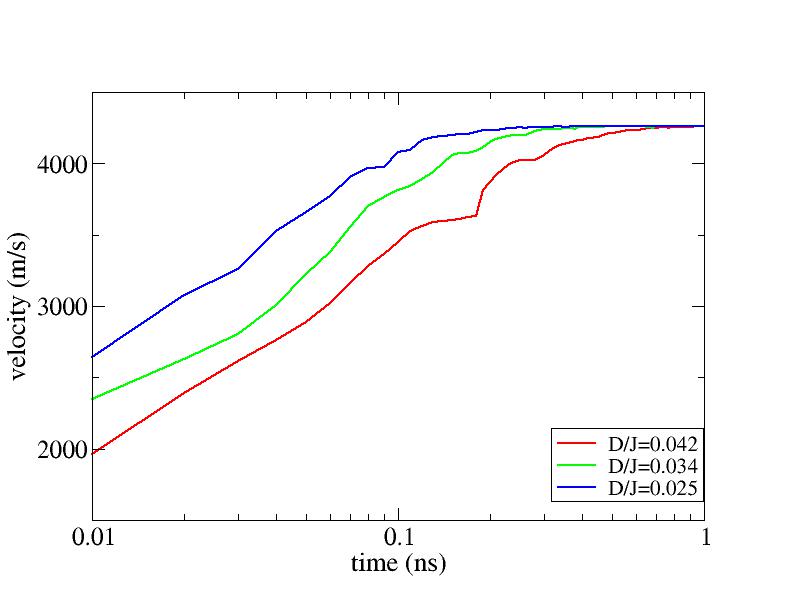}
\caption{\label{sigma_h}\textbf{Velocities as a function of time for three different $D/J$ ratios.} The same parameters were used for all of the systems with $j=0.19\times 10^{12}$ A/m$^2$ and $\beta=0.2$.}
\end{figure}

\vspace{0.5cm}
\noindent{\bf Soliton dynamics and technology.} Soliton dynamics was studied for stable soliton configurations in the presence of  a spin polarised current. In this case, the spin polarised current had a flux of angular momentum which gives rise to a net torque on the soliton that can be driven along the current direction through the spin transfer torque. In the LLG equations we used a current density of $j=0.19\times10^{12}$ A/m$^2$ and the non-adiabatic parameter, which determines the strength of the non-adiabatic torque, was set to be $\beta=2\alpha=0.2$. With these simulations we evaluated the speed of the solitons of Fig.~\ref{lattice}, as a function of time. The relationship between the soliton velocity and time is shown in Fig.~\ref{sigma_h}. Here, we observe that the velocity is increasing very fast during the first 0.2 ns, then it levels off up to 0.4 ns and after that is almost constant until the end of the simulation, at a time of 1 ns. For the same set of parameters, we perform calculations by changing the non-adiabatic parameter $\beta$ ($\beta =0.0$, $\beta = 0.05$, $\beta = 0.1$ and $\beta =0.2$) and investigate how the velocity is influenced by this change. The results we extract from these calculations are presented in Table~\ref{distance} and we observe that the soliton is moving even when $\beta = 0$, in other words it seems that $\beta$ is not essential for moving the soliton, although as Table~\ref{distance} shows, the precise value o fthe velocity depends on $\beta$.

\begin{table}[h]
\begin{center}
\begin{tabular}{|c|c|c|}
  \hline \hline
  Current density ($\times 10^{12}$ A/m$^{2}$) & $\beta$ & velocity (m/s) \\
  \hline \hline
  0.19 & 0.0 & 365 \\
  0.19 & 0.05 & 1191 \\
  0.19 & 0.1 & 2178  \\
  0.19 & 0.2 & 3447 \\
  \hline
  \end{tabular}
\end{center}
\caption{\label{distance} Velocities of the soliton with parameters $J=1.88$ mRy, $D=0.08$ mRy and $D/J=0.042$ by applying spin polarised current with different values of the non-adiabatic parameter $\beta$. The velocities are calculated for $t = 0.1$ ns.}
\end{table}

We performed additional simulations for the time evolution of the spin's texture velocity, tuning the strength of the exchange interaction $J$ and the Dzyaloshinskii-Moriya interaction. In this way, two more systems were examined, the parameters chosen in the first set of simulations were $J=1.96$ mRy and $D=0.05$ mRy with $D/J$ ratio 0.025 and the spin texture is stabilised under the influence of an external magnetic field 0.2 T at zero temperature. The parameters chosen in the second set were  $J=1.91$ mRy, $D=0.065$ mRy and the $D/J$ ratio 0.034, where an external magnetic field of 0.3 T is applied for stabilising the soliton at zero temperature. The time evolution of the velocities are shown in Fig.~\ref{sigma_h} and in both cases the parameters of the spin polarised current are the same as above. The trends of these two last systems are the same as in the first case; the velocity increases very fast for 0.1 ns, then the increase is smoother and the only difference with the first case is that they reache a saturation earlier at approximately 0.5 ns.

Comparing the results obtained with the different choices of $D/J$ ratios we see that initially the dynamics differs a bit between the different systems. It is observed that at the beginning, for $t=0.01$ ns, the velocities are different which implies that an increasing strength of the ferromagnetic nearest-neighbour interaction causes a reduced acceleration of the soliton motion. Eventually, the three different systems obtain a more steady-state motion with a constant velocity. The observed acceleration phase of the solitons is close compared with what is normally observed in regular ferromagnetic domain wall motion \cite{PhysRevLett.92.207203}.

\begin{figure*}
\centering\includegraphics[scale=3.600]{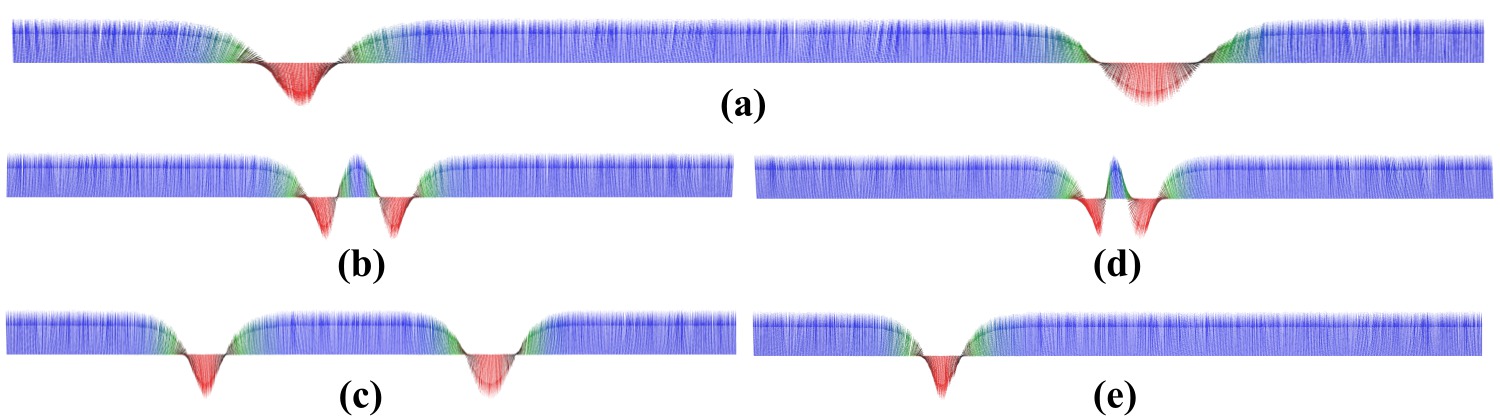}
\caption{\label{two_solitons}\textbf{Collision (b and c) and annihilation (d and e) of two solitons when they are moving in the opposite direction.} \textbf{(a)} The initial positions of the stabilised solitons. \textbf{(b)} Snapshot of the position of the solitons at the moment of their collision when spin polarised current is applied with $j=0.19\times10^{12}$ A/m$^2$ and $\beta=0.2$. \textbf{(c)} Relaxed position of the solitons after the removal of the spin current. \textbf{(d)} Snapshot during the annihilation of solitons after applying spin current with $j=0.38\times10^{12}$ A/m$^2$ and $\beta=0.2$ and \textbf{(e)} creation of one soliton after the annihilation and turn off the spin current.}
\end{figure*}

\begin{figure}
\includegraphics[scale=0.42]{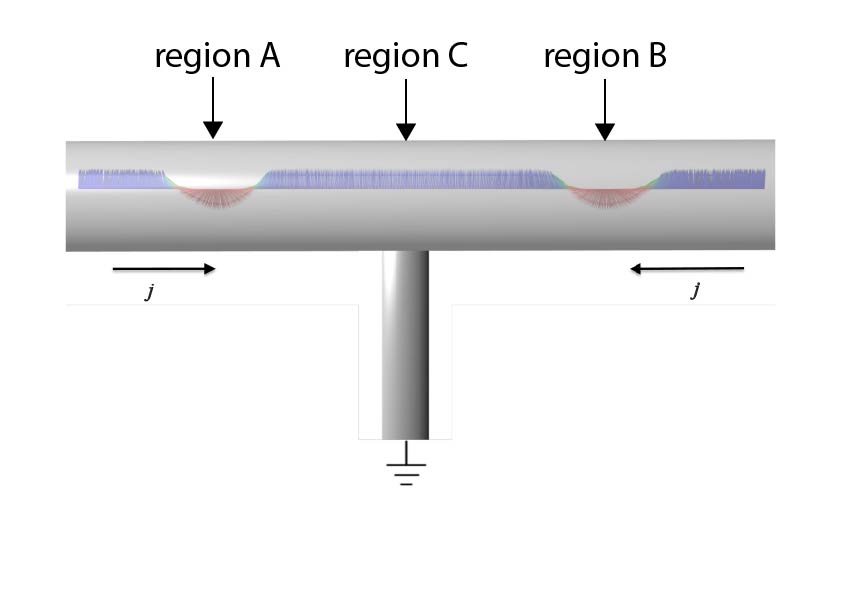}
\caption{\label{experiment}\textbf{Schematic visualisation of a suggestion for applying two opposite currents in experiment.} With the presence of the grounding at the centre of the sample we are able to apply these opposite currents in order to accomplish the collision of solitons, shown in Fig.4 and the supplementary videos. Regions A, B and C discussed in the text.}
\end{figure}

One way of moving these kind of spin textures is to use a spin polarised current. After investigating the current-driven dynamics of isolated solitons  we now turn our attention towards direct manipulation of interacting solitons. 
In our case after the stabilisation of two solitons at the beginning and at the end of our system (Fig.~\ref{two_solitons}a) by applying a local magnetic field (Supplementary Movie 1, is an example of creating one soliton), two spin polarised currents are applied in the system in opposite directions and the solitons start moving towards each other. A suggestion for an experimental setup is shown in Fig.~\ref{experiment}, and initially the two solitons are located in region A and region B). At the moment of their collision (region C), the currents are removed from the system. In the absence of an external driving force, the repulsive interaction between the solitons dominate their dynamics and as a result the solitons move backwards, from each other, and relax after some time (Supplementary Movie 2 and  Fig.~\ref{two_solitons}c). If we apply larger spin currents into the system then the driving force overcomes the repulsive soliton-soliton interactions and as a result the solitons are forced to collide (see Fig.~\ref{two_solitons} d and e). This results in an annihilation mechanism and during the annihilation process, spin waves are emitted (Supplementary Movie 3). At the moment of annihilation the spin currents are turned off and we find that after a while new solitons may emerge. The emergence of new solitons is driven by the previously emitted spin waves which causes small fluctuations of the magnetisation. Since the simulations are performed at a small but finite temperature, stochastic fluctuations are also present in the system and as a result, the position of the newly created soliton/solitons is then not deterministically controllable. However, Fig.~\ref{two_solitons} shows the interesting possibility to control the number of solitons in the system. We started with the presence of two solitons (Fig.~\ref{two_solitons}a) and after applying and removing the two opposite spin currents we end up with only one soliton (Fig.~\ref{two_solitons}e).  

\begin{figure*}
\centering\includegraphics[scale=4.4]{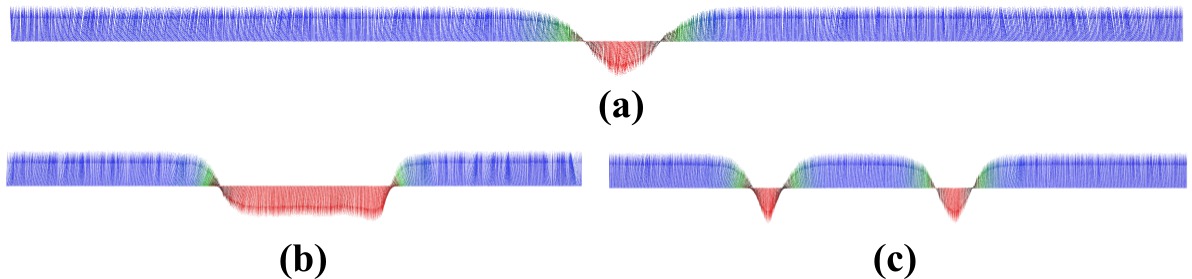}
\caption{\label{experiment2}\textbf{Creation of two solitons by applying a reversed spin polarised current at the centre of one soliton. (a)} The initial position of the stabilised soliton at the centre. \textbf{(b)} Snapshot of the deformation of the soliton at the moment of applying the reversed spin polarised current with $j=6.1\times10^{12}$ A/m$^2$ and $\beta=0.2$. \textbf{(c)} Relaxed position of the solitons after the removal of the spin current.}
\end{figure*}

If now only one soliton is stabilised at the centre of the system (region C, in Fig.\ref{experiment}) and the spin-currents are reversed, we find that a new soliton is created which results in two stable solitons. This process is shown in Fig.~\ref{experiment2} and in Supplementary Movie 4. The reversed currents, with current density $j=6.1\times10^{12}$ A/m\textsuperscript{2} and $\beta=0.2$, are acting on the single soliton (Fig.~\ref{experiment2}a), dragging it in opposite directions (Fig.~\ref{experiment2}b) and this results in the deformation of the soliton. This is caused by the spin transfer torque imposed by the spin-current and the magnetic moments are forced to flip. After very short time we stop the flow of the current and the system minimises its energy and in this process some of the atomic moments are flipping back. The perturbation which is caused by the current is big enough for the system to stabilise two solitons (Fig.~\ref{experiment2}c) when the system is recovering after the removal of the spin current.

A  suggested device for how to realize experimentally soliton collisions that are discussed around Fig.~\ref{two_solitons}, is outlined in Fig.~\ref{experiment}. Other geometries and realizations of the results found in Fig.~\ref{two_solitons} or Fig.~\ref{experiment2} may be possible, but we discuss here some requirements that would be needed to realize an experiment based on the geometry of Fig.~\ref{experiment}. The solitons are initially located at region A and region B. The location of solitons at these regions can be realized by imposing local modifications of the materials parameters in these regions, e.g. by reducing the diameter of the nano-device in this region. This is experimentally realized for domain walls.
Also, from Figs.~\ref{magnfield} and \ref{sigma_h} we note that typical sizes and velocities of the solitons discussed here are in the range of 40-80 nm, respectively 2000-4000 m/s.  A relevant time scale for a successful experimental realization of Fig.~\ref{experiment} would be set by the time a soliton would take before it leaves a region of space equal to its own size. This puts a temporal limit of the electronics needed to control the spin-currents driving the colliding solitons, as discussed in Fig.~\ref{two_solitons}. With the size and speed mentioned above we estimate that electronics in the frequency range 25-100 GHz is needed, which is within current technological abilities \cite{1.4898042}. For the event when one solition, located at region C, turns in to two solitons, by reversal of the current (simulations in Fig. \ref{experiment2}), it is possible to ensure that the initial location of the singular soliton is at region C, by a local modification of the parameters of the nano-device, e.g. by changing the diameter, as discussed above.

We have discussed here possible ways to both annihilate and create solitons with the use of spin polarised currents, in a controlled way, and we have proposed a potential devise which has technological relevance since it easily to create and destroy bits (solitons). A two state device can hence be realised, where the presence of one or two solitons in the device is simply detected by the resistance of the device. Our results are comparable with results of domain wall motion, albeit the soliton solution is shown here to be particularly useful for the device outlined in Fig.~\ref{experiment}. For comparison we note that a micro magnetic study for domain wall motion by spin polarised current in a nanowire \cite{Thiaville} shows that the threshold current density for moving the domain wall is $j=8.5\times10^{12}$A/m\textsuperscript{2} and in our case we are measuring the soliton velocities with current density $j=0.19\times10^{12}$ A/m\textsuperscript{2}. The domain wall in order to reach a velocity of 1400 m/s current density of $j=22.8\times10^{12}$A/m\textsuperscript{2} is needed, on the other hand in our case with much smaller current density we reach velocities of approximately 4200 m/s. Thus, technology proposed in our work has advantage and happens because of soliton in our case propagates through kink crystal in contrast to domain walls which spreads through uniformly polarised ferromagnet. 

In the end of this section we would like to point out that LLG allows a certain class of analytical solutions, e.g., propagating domain wall or Walker's dynamical solution. The latter is known to describe domain wall motion in an infinite medium that is characterized by uniaxial anisotropy \cite{Dillon}. Such a solution was shown to survive under applied uniform magnetic field below a critical value which is purely determined by Gilbert damping $\alpha$ \cite{Walker}. The idea to use propagating domain walls for future spintronics applications looks promising and has been recently addressed from different perspective, e.g., spin waves emission induced domain wall propagation in a magnetic nanowire with easy axis along the wire was studied for a finite transverse magnetic anisotropy \cite{Wang1}. In the meanwhile, it turns out that Walker's solution is not stable with respect to spin waves emission \cite{Wang1,Hu}, making numerical simulations indispensable in studying soliton propagation. Interestingly, spin waves emitted by one domain wall  traveling through another domain wall can lead to attraction between them via spin transfer \cite{Wang1}. In line with our results the authors \cite{Wang2} showed that DM exerts an extra torque which rotates domain wall plane when spin waves pass through it, whereas the presence of the effective easy plane anisotropy suppresses this rotation resulting in a faster domain wall motion.

\section {Discussion}

There basically exist two classes of independent solutions of the sine-Gordon equation, describing nonlinear excitations of spiral magnetic structure. They are stable solitons (extra kinks or breathers) and spin-wave pulses. In the former case, the individual motion of a soliton is always associated with a local translation of a magnetic spiral, as one can observe from our numerical simulations. A direct analytical relation between the soliton's parameters, and local translation, can be used to control solitons or for logical switch operation. In principle, macroscopic translation of solitons in spiral structures might be visualised with magneto-optical measurements or magnetic force microscopy. In the proximity of a commensurate-incommensurate transition, the solitonic motion can be easily controlled by tuning temperature or by an applied external magnetic field, which is perpendicular to the helical axis. In addition, the presence of breathers can be investigated by applying microwave pulses corresponding to eigenfrequencies of the breather. The second class of nonlinear excitations is comprised of spin-wave pulses. The formation of such pulses does not lead to macroscopic translation of the background structure. In fact, magnetisation in propagating spin waves fluctuates around a nonuniform spiral state, which corresponds to an equilibrium magnetic structure. In this case the interaction is not sufficient to stabilise the pulses, and pulses that initially are localised, spread in time thanks to dispersion. Small-amplitude pulses can be approximated with a superposition of linear spin waves corresponding to helical structure.

By combining analytical and numerical work, we have established stability criteria for these magnetic solitons. The parameter space relevant for the soliton solutions is, rewardingly, the same in the analytical and numerical work, and is an internal test of the robustness of our theory. We have in addition shown that it is possible to annihilate and create solitons by means of a spin-polarised current, which couples to the magnetic sub-system via a spin-transfer torque. Concrete suggestions of magnetic devices have been proposed, and it is demonstrated that with experimentally realistic currents, two state solutions can be induced on a time-scale of 325 ps for the annihilation and of 255 ps for the creation. This creates a basis for new technologies in magnetic storage as well as for logical operations using magnetism. Our work hence establishes magnetic solitons as useful technology enablers, and as proposed here, can be useful in the emerging field of solitonics.

\section{Methods}
\noindent{\bf Model Hamiltonian.} The considered one-dimensional lattice is described by the following atomistic Hamiltonian

\begin{equation}\label{latham}
H=-\frac{1}{2}\sum\limits_{i\neq j}J_{ij}\mathbf{m}_i\cdot\mathbf{m}_j-\frac{1}{2}\sum\limits_{i\neq j}\mathbf{D}_{ij}\cdot\left(\mathbf{m}_i\times\mathbf{m}_j\right)-\mathbf{B}\cdot\sum\limits_i\mathbf{m}_i
\end{equation}

\vspace{0.5cm}
\noindent where $i$ and $j$ are atomic indices, $\mathbf{m}_i$ is the classical atomic moment, $J_{ij}$ is the strength of the exchange interaction, $\mathbf{D}_{ij}$ is the anisotropic DM interaction due to spin-orbit coupling, and the last term stands for the external applied magnetic field. It should be noted at this point that the anisotropy term is neglected in the Hamiltonian, due to the fact that this term is the less important compared with the other three terms (exchange interaction, DM interaction and Zeeman term). In the following we use model parameters.

\vspace{0.5cm}
\noindent{\bf Atomistic spin dynamics.} To study the motion of the solitons in 1D structure in the presence of a spin polarised current, we have performed atomistic spin dynamics simulations based on the Landau-Lifshitz-Gilbert (LLG) equation with additional terms to describe the spin transfer torque effect. The resulting equation of motion becomes

\begin{small}
\begin{eqnarray}\nonumber
\frac{\partial\mathbf{m}_i}{\partial t}=-\frac{\gamma}{1+\alpha^2}\mathbf{m}_i\times\left(\mathbf{B}_i^\mathrm{eff}+\frac{\alpha}{m_i}\left(\mathbf{m}_i\times\mathbf{B}_i^\mathrm{eff}\right)\right) \\ \label{LaLi}
+\frac{1+\alpha\beta}{1+\alpha^2}u_x\mathbf{m}_i\times\left(\mathbf{m}_i\times\frac{\partial\mathbf{m}_i}{\partial x}\right)-\frac{\alpha-\beta}{1+\alpha^2}u_x\mathbf{m}_i\times\frac{\partial\mathbf{m}_i}{\partial x}
\end{eqnarray}
\end{small}

\vspace{0.5cm}
\noindent where $\alpha$ is the Gilbert damping coefficient, $\gamma$ the gyromagnetic ratio, $\mathbf{m}_i$ the magnitude of the magnetic moment, $\mathbf{B}_i^\mathrm{eff}$ the effective field resulting from the spin-Hamiltonian in equation (\ref{latham}), \noindent $\beta$ the so-called non-adiabatic parameter and $u_x$ the velocity term. The resulting effective field is given by:

\begin{equation}
\mathbf{B}_i^\mathrm{eff}=\mathbf{B}_i+\mathbf{b}_i(t)
\end{equation}

\vspace{0.5cm}
\noindent where $\mathbf{B}_i=-\partial H/\partial\mathbf{m}_i$ and $H$ refers to the Hamiltonian (\ref{latham}). The temperature effects on the spin system are taken into account through Langevin dynamics; the thermal fluctuations are represented here by $\mathbf{b}_i(t)$ which can be modelled by Gaussian correlators

\begin{equation}
\langle b_i^\mu(t)\rangle=0,\; \langle b_i^\mu(t)b_j^\nu(t^\prime)\rangle=\frac{2\alpha_ik_BT\delta_{ij}\delta_{\mu\nu}\delta(t-t^\prime)}{\gamma m_0(1+\alpha_i)^2}
\end{equation}

\vspace{0.5cm}
\noindent where $\mu$, $\nu$ are the indexes of Cartesian coordinates.

The velocity term is in units of velocity (m/s) and is related to the magnitude and the direction of the current $\mathbf{j}$, is proportional to the applied current and is equal to:

\begin{equation}
u_x=\frac{g\mu_BPj}{2eM_s}
\end{equation}

\vspace{0.5cm}
\noindent where $j$ is the current density, $P$ the spin polarisation of the current and $M_s$ the saturation magnetisation per unit cell.

\section{Acknowledgements}
We acknowledge support from the KAW foundation (projects KAW 2013.0020 and KAW 2012.0031) as well as the Swedish Research Council (VR) and eSSENCE. Calculations performed with support from SNIC/SNAC. D.Y acknowledges the support from RFBR project 16- 32-60040. The authors acknowledge useful comments from the two referees of this paper.

\section{Author contributions}
D.Y designed the theoretical model and performed analytical, theoretical analysis. K.K. performed the atomistic spin dynamics calculations. A.B., D.Y., O.E. and K.K. analysed the data. K.K. and D.Y. co-wrote the manuscript. All authors discussed the data and commented on the manuscript.

\section{Competing Interests}
The authors declare that they have no competing financial interests.

\end{document}